\documentclass[journal=jctcce,manuscript=article,layout=traditional]{achemso}
\usepackage[version=3]{mhchem} 
\usepackage[table,xcdraw,dvipsnames]{xcolor}
\usepackage{fullpage}
\usepackage{mathtools}
\usepackage{hyperref,url}
\usepackage{amsmath}
\usepackage{amsfonts}
\usepackage[capitalise]{cleveref}
\usepackage{amsthm}
\usepackage{amssymb}
\usepackage{bm}
\usepackage{braket}
\usepackage{relsize}
\usepackage{enumitem}
\usepackage[font=small,labelfont=bf]{caption}
\usepackage{graphicx}
\usepackage{footnote}
\usepackage{multirow}
\usepackage{tabularx}
\usepackage{booktabs}
\usepackage{makecell}
\usepackage{threeparttable}
\usepackage{chemfig}
\usepackage{subcaption}
\usepackage{physics}
\usepackage{comment}
\usepackage{algpseudocode}
\usepackage{algorithm}

\usepackage{lipsum}
\usepackage{epstopdf}

\author{Adam Rettig}
\affiliation{Department of Chemistry, University of California, Berkeley, California 94720, USA}
\alsoaffiliation{Chemical Sciences Division, Lawrence Berkeley National Laboratory, Berkeley, California 94720, USA}

\author{Joonho Lee}
\affiliation{Department of Chemistry, Columbia University, New York, New York 10027, USA}
\email{joonholee@g.harvard.edu}

\author{Martin Head-Gordon}
\affiliation{Department of Chemistry, University of California, Berkeley, California 94720, USA}
\alsoaffiliation{Chemical Sciences Division, Lawrence Berkeley National Laboratory, Berkeley, California 94720, USA}
\email{mhg@cchem.berkeley.edu}

\title {Even Faster Exact Exchange for Solids via Tensor Hypercontraction}

\abbreviations{PBC, K, THC}

\begin{document}

%
%
\begin{abstract}
Hybrid density functional theory (DFT) remains intractable for large periodic systems due to the demanding computational cost of exact exchange. We apply the tensor hypercontraction (THC) (or interpolative separable density fitting) approximation to periodic hybrid DFT calculations with Gaussian-type orbitals. This is done to lower the computational scaling with respect to the number of basis functions  ($N$), and $\mathbf k$-points ($N_k$). Additionally, we propose an algorithm to fit only occupied orbital products via THC (i.e. a set of points, $N_\text{ISDF}$) to further reduce computation time and memory usage. This algorithm has linear scaling cost with $\mathbf k$-points, no explicit dependence of $N_\text{ISDF}$ on basis set size, and overall cubic scaling with unit cell size.  Significant speedups and reduced memory usage may be obtained for moderately sized systems, with additional gains for large systems. Adequate accuracy can be obtained using THC-oo-K for self-consistent calculations. We perform illustrative hybrid density function theory calculations on the benzene crystal in the basis set and thermodynamic limits to highlight the utility of this algorithm.
\end{abstract}

%
%
\section{Introduction}
Density functional theory (DFT) has dominated the field of quantum chemistry for the last several decades due to its combination of satisfactory accuracy and relatively low $\mathcal O(N^3-N^5)$ computational cost, where $N$ is the number of basis functions. Within density functional theory, hybrid functionals, which include Hartree-Fock ``exact exchange'' (shown in \cref{eq:exchange}),
\begin{equation}
    K_{\mu\nu} = \sum_{\lambda \sigma} (\mu \lambda | \sigma \nu) P_{\lambda \sigma}
    \label{eq:exchange}
\end{equation}
have emerged as the most accurate and widely used functionals in nearly all applications for molecules \cite{Mardirossian:2017b,Goerigk:2017,Dohm:2018, Chan:2019}, with only a mild increase in the computational cost compared to local density functionals. Unsurprisingly, hybrid density functional theory accounts for a large portion of quantum chemical studies performed on molecules today.

Hybrid functionals are known to offer large improvements in accuracy for some systems under periodic boundary conditions (PBC) as well\cite{muscat2001prediction,batista2006comparison, deak2010accurate, komsa2011assessing, henderson2011accurate}. Despite this, local (or pure) density functionals account for the majority of PBC studies due to their computational efficiency. Periodic calculations are often significantly more expensive than typical molecular studies because one needs to reach the thermodynamic limit (TDL). 
By virtue of translational symmetry, one may reach TDL results by numerically integrating over the first Brillouin zone (i,e, by sampling $\mathbf k$-points). This approach leads to lower scaling than equivalent supercell approaches. In local DFT, different $\mathbf k$ points can be treated independently, leading to an extra factor in the computational scaling equal to the number of $\mathbf k$ points, $N_k$. Hybrid density functional theory incurs an extra penalty as the different $\mathbf k$ points are coupled through the periodic exact exchange, shown in \cref{eq:pbcexchange}.
\begin{equation}
    K_{\mu\nu}^k = \sum_{\lambda \sigma k'} (\phi_\mu^k \phi_\lambda^{k'} | \phi_\sigma^{k'} \phi_\nu^k) P_{\lambda \sigma}^{k'}.
    \label{eq:pbcexchange}
\end{equation}
This leads to an extra computational scaling factor of $N_k^2$, giving an overall computational scaling for periodic exact exchange of $\mathcal O(N_k^2 N^3)$ compared to the $\mathcal O(N_k N^3)$ of local functionals. $\mathbf k$-point calculations yield a massive reduction in computational cost for both local and hybrid functionals compared to equivalent supercell calculations that would scale as $\mathcal O(N_k^3 N^3)$ in this notation. Despite the computational speedup of $\mathbf k$-point calculations, hybrid functionals remain considerably more expensive than local functionals for periodic systems with an additional overhead of $N_k$. Furthermore, hybrid functionals converge slower to the TDL than local functionals, necessitating even larger $k$-point calculations, further adding to the computational cost.

Efficient exact exchange computation for molecular systems has been the subject of considerable study over the last few decades; many of these ideas have also been transferred to periodic systems. For sufficiently large systems, the number of nonzero elements in the two-electron integral tensor increases only quadratically with the system size, motivating sparsity-aware approaches to lower the scaling of exact exchange to $\mathcal O(N)$ for gapped systems \cite{schwegler1996linear, schwegler1997linear, burant1996linear, challacombe1997linear,ochsenfeld1998linear}. In practice, this asymptotic region is hard to reach, limiting the usefulness of such approaches. Approximate factorization of the two-electron integral tensors into lower rank tensors has proved more successful, leading to the pseudospectral method \cite{friesner1985solution} (and the closely related chain of spheres method\cite{neese2009efficient, izsak2011overlap}), Cholesky decomposition\cite{beebe1977simplifications, roeggen1986beebe}, the resolution of the identity (RI) or density fitting approximation \cite{baerends1973self,whitten1973coulombic,jafri1974electron}, and tensor hypercontraction (THC)\cite{hohenstein2012tensor,parrish2012tensor}. Of these, RI, has emerged as the most popular in both exact exchange and correlated wavefunction theory due to its large speedup and its relatively robust, low error of $\sim 50 \mu$E$_H$ per atom (which also often cancels in energy differences).

Many periodic codes utilize plane waves as a basis set in contrast to the Gaussian-type orbitals (GTO) used in the molecular case, complicating the direct application of these algorithms to periodic calculations. The adaptively compressed exchange (ACE) algorithm, similar to the occ-RI-K algorithm developed for molecular systems, has become the most common way to speed up exact exchange evaluation in planewave-based periodic quantum chemistry codes\cite{lin2016adaptively}. More recently, the interpolative separable density fitting (ISDF)\cite{shenvi2014tensor,lee2019systematically,dong2018interpolative} approach to THC has been shown to offer considerable speedup for both gamma point and $\mathbf{k}$-point plane wave calculations\cite{hu2017interpolative, wu2021low, li2022complex, qin2023interpolative}.

We use GTOs for periodic calculations via the Bloch orbital framework, allowing for a more direct application of molecular-based exact exchange algorithms. Our crystalline GTO basis functions are given by:
\begin{equation}
    \phi_\mu^{\mathbf{k}}(\mathbf{r}) = \frac{1}{\sqrt{N_k}} \sum_{\mathbf{R}} e^{i\mathbf{k} \cdot \mathbf{R}} \tilde{\phi}_\mu(\mathbf{r} - \mathbf{R}) =  \frac{1}{\sqrt{N_k}} e^{i\mathbf{k} \cdot \mathbf{r}} u^{\mathbf{k}}_\mu(\mathbf{r}) 
\end{equation}
where $\mathbf{R}$ are direct lattice vectors, $\mathbf{k}$ is the crystalline momentum, and $\tilde{\phi}_\mu(\mathbf{r})$ is an atomic orbital. $u^{\mathbf{k}}_\mu(\mathbf{r})$ is the cell periodic part of the Bloch orbital, which is periodic along the lattice (i.e., $u^{\mathbf{k}}_\mu(\mathbf{r}) = u^{\mathbf{k}}_\mu(\mathbf{r} + \mathbf{R})$), and is given by:
\begin{equation}
     u^{\mathbf{k}}_\mu(\mathbf{r}) = \frac{1}{\sqrt{N_k}} \sum_{\mathbf{R}} e^{i\mathbf{k} \cdot (\mathbf{R} - \mathbf{r})} \tilde{\phi}_\mu(\mathbf{r} - \mathbf{R})
\end{equation}

Recently, our group showed the extension of the molecular occ-RI-K algorithm\cite{Manzer:2015b} to GTO-based PBC calculations leads to significant speedups for large basis sets, up to two orders of magnitude for the systems studied\cite{lee2022faster}. ISDF approaches have also been developed for both GTO and numerical atomic orbital (NAO) based $\Gamma$-point PBC calculations \cite{sharma2022fast,qin2020interpolative}. 

In this study, we propose a THC algorithm for periodic exact exchange with $\mathbf k$-point sampling in a GTO basis utilizing ISDF. Similar to the analogous plane wave implementation\cite{wu2021low}, the use of ISDF can reduce the computational scaling of exact exchange with respect to the number of $\mathbf k$-points. We will additionally propose a new ISDF approach by fitting only products of occupied orbitals, as is done in ACE-ISDF in plane waves to realize further computational savings. We illustrate the scaling improvements of these algorithms and perform an illustrative hybrid DFT study of the benzene crystal cohesive energy.

\section{Theory}

\subsection{Review of Molecular and $\Gamma$-point ISDF}
In molecular and $\Gamma$-point periodic ISDF, the products of atomic orbitals are approximated by a sum of interpolation vectors $\{\xi_P(r)\}$ weighted by the orbitals evaluated at a set of $N_\text{ISDF}$ interpolation points $\{r_P\}$:
\begin{equation}
    \phi_\mu(\mathbf{r})^* \phi_\nu(\mathbf{r}) \approx \sum_P^{N_\text{ISDF}} \phi_\mu(\mathbf{r}_P)^* \phi_\nu(\mathbf{r}_P) \xi^{[nn]}_P(\mathbf{r})
\end{equation}
In the above, we have labeled the interpolation vectors with the $[nn]$ superscript to denote that these functions are fit to products of two atomic orbitals. 

The set of interpolation points $\{\mathbf r_P\}$ can be selected in numerous ways; commonly, either a QR decomposition with column pivoting\cite{hu2017interpolative} or a centroidal Voronoi tessellation (CVT) k-means algorithm\cite{dong2018interpolative} is used. The number of interpolation points is set via a parameter $c_\text{ISDF}^{[nn]}$:
\begin{equation}
N_\text{ISDF}^{[nn]} = c_\text{ISDF}^{[nn]} N.
\label{eq:N_ISDF^nn}
\end{equation}
After the interpolation points are chosen, the interpolation vectors $\xi^{[nn]}_P(r)$ are determined via a least squares fit\cite{parrish2012tensor, matthews2020improved} to the orbital products desired:
\begin{equation}
    \sum_Q \mathbf S^{[nn]}_{PQ} \xi_Q^{[nn]}(\mathbf{r}) = \sum_{\mu \nu}  \phi_\mu(\mathbf{r})^* \phi_\nu(\mathbf{r})  \phi_\nu(\mathbf{r}_P)^* \phi_\mu(\mathbf{r}_P)
    \label{eq:interpvec}
\end{equation}
In the above, $\mathbf S$ represents the ISDF metric matrix (not to be confused with the AO overlap matrix):
\begin{equation}
    S_{PQ}^{[nn]} = \sum_{\mu\nu} \phi_\mu (\mathbf{r}_P)^* \phi_\mu (\mathbf{r}_Q)^*  \phi_\nu (\mathbf{r}_P)^*  \phi_\nu (\mathbf{r}_Q)
\end{equation}

The exchange matrix may then be computed using only the basis functions evaluated at a set of interpolation points and the two-electron integrals between interpolation vectors, $M_{PQ}$, of which there are only $N_\text{ISDF}^2$. 
\begin{equation}
    M_{PQ} = \int\int d\mathbf{r}_1 d\mathbf{r}_2 \frac{\xi_P(\mathbf{r}_1) \xi_Q(\mathbf{r}_2) }{r_{12}}
    \label{eq:M}
\end{equation}
$M_{PQ}$, given in \cref{eq:M} may be computed in reciprocal space according to the GPW algorithm. The cost of computing $M_{PQ}$ is ${\mathcal O}(N_g N_\text{ISDF}^2)$ due to the contraction of the two interpolation vectors in reciprocal space.

$M_{PQ}$ can be very expensive to compute for large systems, but needs only be computed once at the start of a calculation. It may then be used each iteration to compute $K_{\mu\nu}$ according to \cref{eq:thcexchange}.
\begin{equation}
    K_{\mu\nu} = \sum_{\lambda \sigma} \phi_\mu(\mathbf{r}_P)^* \phi_\lambda(\mathbf{r}_P) M_{PQ} \phi_\sigma(\mathbf{r}_Q)^* \phi_\nu(\mathbf{r}_Q) P_{\lambda \sigma}
    \label{eq:thcexchange}
\end{equation}
This is done most efficiently by first contracting $\phi_\lambda(\mathbf{r}_P) \phi_\sigma(\mathbf{r}_Q)^* P_{\lambda \sigma}$ and taking a Hadamard (element-wise) product with $M_{PQ}$. This intermediate is then contracted with $\phi_\mu(\mathbf{r}_P)^*$ and then $\phi_\nu(\mathbf{r}_Q)$, which is the bottleneck at O$(N N_\text{ISDF}^2)$ cost. As $N << N_g$, the per iteration cost is significantly less than the initial formation of $M_{PQ}$. Sharma et al. showed that large speedups were possible with this algorithm for periodic $\Gamma$-point calculations\cite{sharma2022fast}.

\subsection{Periodic $\mathbf k$-point ISDF}
We first examine the extension of the $\Gamma$-point algorithm presented above to $\mathbf k$-point GTO-based calculations. We do this by fitting the ISDF interpolation vectors to only the cell periodic part of our basis functions, $u^{\mathbf{k}}_{\mu}(r) $\cite{wu2021low}:
\begin{equation}
    u^{\mathbf{k}}_{\mu}(\mathbf{r})^* u^{\mathbf{k}'}_{\nu}(\mathbf{r}) \approx \sum_P u^{\mathbf{k}}_{\mu}(\mathbf{r}_P)^* u^{\mathbf{k}'}_{\nu}(\mathbf{r}_P) \xi^{[nn]}_P(\mathbf{r})
    \label{eq:kpointinterpfunc}
\end{equation}
The full product of basis functions may then be approximated by adding in the Bloch phase factors:
\begin{equation}
    \phi_{\mu}^{\mathbf{k}}(\mathbf{r})^* \phi_{\nu}^{\mathbf{k}'}(\mathbf{r}) \approx \sum_P u_{\mu}^{\mathbf{k}}(\mathbf{r}_P)^* u_{\nu}^{\mathbf{k}'}(\mathbf{r}_P) e^{i\mathbf{q}\cdot \mathbf{r}}\xi^{[nn]}_P(\mathbf{r})
    \label{eq:kpointinterpfunc}
\end{equation}
In the above we have defined $\mathbf{q}=\mathbf{k}'-\mathbf{k}$. Note that our interpolation vectors do not depend on $\mathbf k$-points. 

The interpolation functions are once again determined via a least squares fit, but we now must include a sum over $\mathbf k$-points:
\begin{equation}
    \sum_{Q}\mathbf S^{[nn]}_{PQ} \xi_Q^{[nn]}(\mathbf{r}) = \sum_{\mu \nu \mathbf{k} \mathbf{k}' } u_\mu^{\mathbf{k}}(\mathbf{r})^* u_\nu^{\mathbf{k}'}(\mathbf{r})  u_\nu^{\mathbf{k}'}(\mathbf{r}_P)^* u_\mu^{\mathbf{k}}(\mathbf{r}_P)
    \label{eq:pbcinterpvec}
\end{equation}
The $\mathbf{S}$ matrix will similarly include a sum over $\mathbf k$ points. The $\mathbf k$-point $\mathbf{K}$ matrix may then be computed as:
\begin{equation}
    K_{\mu\nu}^{\mathbf{k}} = \sum_{\lambda \sigma \mathbf{q} P Q} u_\mu^{\mathbf{k}}(\mathbf{r}_P)^* u_\lambda^{\mathbf{k}+\mathbf{q}}(\mathbf{r}_P) M_{PQ}^{\mathbf{q}} u_\sigma^{\mathbf{k}+\mathbf{q}}(\mathbf{r}_Q)^* u_{\nu}^{\mathbf{k}}(\mathbf{r}_Q) P_{\lambda \sigma}^{\mathbf{k}+\mathbf{q}}
    \label{eq:kpointthck}
\end{equation}
We have folded the Bloch phase factors included in our basis functions into our definition of $\mathbf{M}$, so that it now has a $\mathbf{q}$-dependence:
\begin{equation}
    M^{\mathbf{q}}_{PQ} = \int\int d\mathbf{r} d\mathbf{r}' \frac{e^{i\mathbf{q}\cdot\mathbf{r}_1} \xi_P(\mathbf{r}_1)^* e^{-i\mathbf{q}\cdot\mathbf{r}_2} \xi_Q(\mathbf{r}_2) }{r_{12}}
\end{equation}
However, $\mathbf{M}$ depends only on $\mathbf{q}$, rather than both $\mathbf{k}$ and $\mathbf{k'}$. The number of unique two-electron integrals is, therefore, linear in the number of $\mathbf{k}$-points. $\mathbf{M}$ can be computed via GPW in O($N_k N_\text{ISDF}^2 N_g$) time. However, this intermediate may be computed once, and stored for reuse in each iteration.

Naively, evaluating \cref{eq:kpointthck} would lead to O($N_k^2 N N_\text{ISDF}^2 $) scaling. However, we write $\mathbf{k}'$ as $\mathbf{k} + \mathbf{q}$ to emphasize that the sum over $\mathbf{q}$ done in the contraction of $M^{\mathbf{q}}_{PQ}$ with $u_\lambda^{\mathbf{k}+\mathbf{q}}(\mathbf{r}_P)u_\sigma^{\mathbf{k}+\mathbf{q}}(\mathbf{r}_Q)^*P_{\lambda \sigma}^{\mathbf{k}+\mathbf{q}}$ is  a convolution in $\mathbf k$ space. Convolutions may be done utilizing FFT in O$(N_k \text{log}(N_k))$ time. Using this trick, we can lower the computational scaling of this algorithm, which we term THC-AO-K, to O($N_k \text{log}(N_k) N N_\text{ISDF}^2 $) per iteration. The rate-limiting step of this algorithm is therefore computing $\mathbf{M}$ due to the large value of $N_g$. In this way, significant savings can be realized for large $\mathbf k$-point calculations - the exchange may be computed only cubically in system size and linearly in $N_{\mathbf k}$. A summary of the THC-AO-K algorithm is shown in Algorithm \ref{alg:thcaok}.

While this algorithm is overall cubic scaling, a significant prefactor is associated with computing the $\mathbf{M}$ matrix. This prefactor was so significant in molecular cases that the THC algorithm does not become competitive with higher scaling approaches unless extremely large systems are studied.\cite{lee2019systematically} In this study, we will analyze the timing of this algorithm to see if the added benefit of reduced $\mathbf k$-point scaling will lead to a crossover at useful system sizes. Additionally, one must be concerned with memory usage, as storing the $\mathbf{M}$ matrix requires O($N_{\text{ISDF}}^2 N_k$) space; this can quickly become impractical to store for large systems and large basis sets.

\begin{algorithm}
\caption{THC-AO-K algorithm}
\label{alg:thcaok}

\begin{algorithmic}
\State $\{ \mathbf{r}_P \} \leftarrow$ CVT k-means
\State Compute $\xi^{[nn]}_P(\mathbf{r})$
\State $\xi^{[nn]}_P(\mathbf{G}) \leftarrow \text{FFT}(\xi^{[nn]}_P(\mathbf{r}))$
\State $M^{\mathbf{q}}_{PQ} \leftarrow \sum\limits_{\mathbf{G}} \xi^{[nn]}_P(\mathbf{G}) V^{\mathbf{q}}(\mathbf{G}) \xi^{[nn]}_Q(\mathbf{G})$
\While{SCF unconverged}
    \State $X^{\mathbf{k}}_{PQ} \leftarrow \sum\limits_{i\mathbf{q}} u_i^{\mathbf{k}+\mathbf{q}}(\mathbf{r}_P) M^{\mathbf{q}}_{PQ} u_i^{\mathbf{k}+\mathbf{q}}(\mathbf{r}_Q)^*$ (FFT Convolve)
    \State $K_{\mu\nu}^{\mathbf{k}} \leftarrow \sum\limits_{PQ} u_\mu^{\mathbf{k}}(\mathbf{r}_P)^* X^{\mathbf{k}}_{PQ} u_\nu^{\mathbf{k}}(\mathbf{r}_Q)$
\EndWhile
\end{algorithmic}

\end{algorithm}

\subsection{THC-oo-K}
Further reduction in compute time and storage requirements can be realized by fitting only products of occupied orbitals via ISDF as seen in \cref{eq:orbprodoo}:
\begin{equation}
    u_{i}^{\mathbf{k}}(\mathbf{r})^* u_{j}^{\mathbf{k}'}(\mathbf{r}) \approx \sum_P u_{i}^{\mathbf{k}}(\mathbf{r}_P)^* u_{j}^{\mathbf{k}'}(\mathbf{r}_P) \xi^{[oo]}_P(\mathbf{r})
    \label{eq:orbprodoo}
\end{equation}
Accordingly, we label the interpolation vectors with the $[oo]$ superscript to indicate that these functions are fit to products of two occupied orbitals. This approach is similar to what is done with plane-wave ISDF codes using the ACE procedure in an algorithm we term THC-oo-K. THC-oo-K is motivated by the fact that the exchange \textit{energy} depends only on occupied orbitals. Of course, the orbital gradient contributions due to exact exchange depend on the virtual orbitals, but they can be computed by directly differentiating the THC-oo-K energy function. In THC-oo-K, the interpolation vectors fit far fewer quantities - which scale only with the number of occupied orbitals (i.e., independent of basis size!). We therefore define the number of ISDF points similarly to Eq. \ref{eq:N_ISDF^nn} via:
\begin{equation}
N_\text{ISDF}^{[oo]} = c_\text{ISDF}^{[oo]} N_{\text{occ}},
\label{eq:N_ISDF^oo}
\end{equation}
with the expectation that $N_\text{ISDF}^{[oo]}  << N_\text{ISDF}^{[nn]} $ (where $nn$ refer to atomic orbitals). The ISDF exchange energy is then computed as before via:
\begin{equation}
    E_X = -\sum_{i j \mathbf{k} \mathbf{q} P Q} u_i^{\mathbf{k}}(\mathbf{r}_P)^* u_j^{\mathbf{k}+\mathbf{q}}(\mathbf{r}_P) M^{\mathbf{q}}_{PQ} u_j^{\mathbf{k}+\mathbf{q}}(\mathbf{r}_Q)^* u_{i}^{\mathbf{k}}(\mathbf{r}_Q) 
    \label{eq:thcoccrik}
\end{equation}

While the potential memory and computational savings arising from the decrease in $N_\text{ISDF}$ seem very promising at first glance, there are several complications that arise from this approach. First, the occupied orbitals change each iteration, necessitating the recalculation of $\xi_P(r)$ and therefore $M_{PQ}^{\mathbf{q}}$ on each iteration. Second, the orbital gradient of the exchange energy is no longer as simple as the THC-AO-K case seen in \cref{eq:thcexchange} due to the dependence of $M$ on the occupied orbitals. The first point shows that there will be a tradeoff in compute time - the computation of $M$ is much cheaper due to the decrease in $N_\text{ISDF}$. However, one must compute $M$ each iteration compared to only once for THC-AO-K. We will subsequently investigate whether this tradeoff is favorable for THC-oo-K in normal use cases. 

The second point can be addressed by deriving the analytical expression for the $vo$ block of the exchange matrix, given by:
\begin{equation}
\begin{aligned}
    K^{\mathbf{k}}_{ai} =&  \sum_{j \mathbf{k}_2 PQ} u_a^{\mathbf{k}}(\mathbf{r}_P)^* u_j^{\mathbf{k}_2}(\mathbf{r}_P) M^{\mathbf{q}}_{PQ} u_j^{\mathbf{k}_2}(\mathbf{r}_Q)^* u_i^{\mathbf{k}}(\mathbf{r}_Q)  \\
    +& \sum_{j \mathbf{k}_2 \mathbf{q} PQR \mathbf{r}}  W_{PQ}^{\mathbf{q}} V^{\mathbf{q}}_P(\mathbf{r}) S^{-1}_{QR} \bigg[ u_a^{\mathbf{k}}(\mathbf{r}_R)^*  u_i^{\mathbf{k}}(\mathbf{r})  u_j^{\mathbf{k}_2}(\mathbf{r})^* u_j^{\mathbf{k}_2}(\mathbf{r}_R) + u_a^{\mathbf{k}}(\mathbf{r})^*  u_i^{\mathbf{k}}(\mathbf{r}_R)  u_j^{\mathbf{k}_2}(\mathbf{r}_R)^* u_j^{\mathbf{k}_2}(\mathbf{r})\bigg]  \\
    -&  \sum_{j \mathbf{k}_2 \mathbf{q} PQST}  W_{PQ}^{\mathbf{q}}  S^{-1}_{QS} M_{PT}^{\mathbf{q}}  \bigg[u_a^{\mathbf{k}}(\mathbf{r}_S)^* u_i^{\mathbf{k}}(\mathbf{r}_T)  u_j^{\mathbf{k}_2}(\mathbf{r}_S)^* u_j^{\mathbf{k}_2}(\mathbf{r}_T) + u_a^{\mathbf{k}}(\mathbf{r}_T)^*  u_i^{\mathbf{k}}(\mathbf{r}_S)  u_j^{\mathbf{k}_2}(\mathbf{r}_T)^* u_j^{\mathbf{k}_2}(\mathbf{r}_S)\bigg]  \\
\end{aligned}
\end{equation}
In the above equation, $W_{PQ}^{\mathbf{q}} = \sum\limits_{ij\mathbf{k}} u_i^{\mathbf{k}}(\mathbf{r}_P)^* u_j^{\mathbf{k}+\mathbf{q}}(\mathbf{r}_P) u_j^{\mathbf{k}+\mathbf{q}}(\mathbf{r}_Q)^* u_i^{\mathbf{k}}(\mathbf{r}_Q) $ where the convolution over $\mathbf k$-points may be done using FFT once again. The dependence of $\mathbf{M}$ on the occupied orbitals leads to additional two terms in the $\mathbf{K}$ matrix, but they are relatively easy to compute as most pieces are needed during the computation of the first term anyway. The effect of these extra terms is an additional prefactor of $\sim3$ in the computation time in our experience. These terms may be contracted in a similar manner to THC-AO-K, leading to an overall scaling of O($N_k N_{\text{ISDF}}^2 N_g$) once again. A summary of the THC-oo-K algorithm is given in Algorithm \ref{alg:thcook}.

In this study, we will investigate the accuracy of both THC-AO-K and THC-oo-K as a function of $N_{\text{ISDF}}$ and analyze their relative computation times across a variety of system sizes.

\begin{algorithm}
\caption{THC-oo-K algorithm}
\label{alg:thcook}

\begin{algorithmic}
\State $\{ \mathbf{r}_P \} \leftarrow$ CVT k-means
\While{SCF unconverged}
    \State Compute $\xi^{[oo]}_P(\mathbf{r})$
    \State $\xi^{[oo]}_P(\mathbf{G}) \leftarrow \text{FFT}(\xi^{[oo]}_P(\mathbf{r}))$
    \State $M^{\mathbf{q}}_{PQ} \leftarrow \sum\limits_{\mathbf{G}} \xi^{[oo]}_P(\mathbf{G}) V^{\mathbf{q}}(\mathbf{G}) \xi^{[oo]}_{\mathbf{Q}}(\mathbf{G})$
    \State $X^{\mathbf{k}}_{PQ} \leftarrow \sum\limits_{iq} u_i^{\mathbf{k}+\mathbf{q}}(\mathbf{r}_P) M^{\mathbf{q}}_{PQ} u_i^{\mathbf{k}+\mathbf{q}}(\mathbf{r}_Q)^*$ (FFT Convolve)
    \State Compute $\frac{\partial M}{\partial \theta^{\mathbf{k}}_{ia}}$
    \State $K_{i j}^{\mathbf{k}} \leftarrow \sum\limits_{PQ} u_i^{\mathbf{k}}(\mathbf{r}_P)^* X^{\mathbf{k}}_{PQ} u_j^{\mathbf{k}}(\mathbf{r}_Q)$
    \State $K_{i a}^{\mathbf{k}} \leftarrow \sum\limits_{PQ} u_i^{\mathbf{k}}(\mathbf{r}_P)^* X^{\mathbf{k}}_{PQ} u_a^{\mathbf{k}}(\mathbf{r}_Q) + \frac{\partial M}{\partial \theta^{\mathbf{k}}_{ia}}$
\EndWhile
\end{algorithmic}

\end{algorithm}

%
%
\section{Computational Details}
\label{compute-details}
Calculations were performed using the  SZV-GTH, DZVP-GTH, TZV2P-GTH, QZV2P-GTH, unc-def2-QZVP\cite{Lee2021Oct} basis sets designed for use in periodic systems with the Goedecker, Teter, Hutter (GTH) pseudopotentials. The GTH-PBE pseudopotential \cite{goedecker1996separable, krack2005pseudopotentials} was used for all DFT calculations, while the GTH-HF pseudopotential was used for all HF calculations.

All calculations were performed using the Gaussian Plane Wave (GPW) algorithm\cite{lippert1997hybrid,vandevondele2005quickstep} for computing 2-electron integrals. A sufficient kinetic energy cutoff for the auxiliary basis was used to fully converge all computations below $10^{-6}$ Hartrees. To treat the divergence of the exact exchange, we use a Madelung correction\cite{fraser1996finite}.

Calculations for the benzene system were performed with a counterpoise correction including basis functions for a (2,2,2) supercell. Cohesive energies for varying $\mathbf k$-point mesh sizes and extrapolated to the thermodynamic limit via:
\begin{equation}
E_\text{TDL} = \frac{E_{N_{k_2}}N_{k_1}^{-1}  - E_{N_{k_1}}N_{k_2}^{-1}}{N_{k_1}^{-1} - N_{k_2}^{-1}}
\end{equation}
All calculations were performed in the Q-Chem software package\cite{epifanovsky2021software,lee2022faster}.

%
%
\section{Results and Discussion}

\subsection{THC Accuracy}

We first investigate the accuracy of both THC algorithms to determine acceptable values of $c_\text{ISDF}$. In the molecular case, it was found that $c^{[nn]}_{\text{ISDF}} = 20-40$ was sufficient to reproduce the RI-level of accuracy of 50 $\mu$E$_H$ per atom for several datasets including thermochemistry and noncovalent interactions\cite{lee2019systematically}. Such a large $c^{[nn]}_{\text{ISDF}}$ was needed to avoid non-variational collapse to spurious solutions (found with smaller $c^{[nn]}_{\text{ISDF}}$). While we did not observe any non-variational collapse in PBC calculations, we found that a similar value is needed for $c^{[nn]}_{\text{ISDF}}$ as we will see. The interpolation vectors must now represent orbital pairs at multiple $\mathbf k$-points however so we expect this value to increase slightly as more $\mathbf k$-points are used, although it should eventually plateau.\cite{wu2021low} Additionally, we seek to determine an appropriate value for $c^{[oo]}_{\text{ISDF}}$, so that the efficiency of THC-AO-K and THC-oo-K can be meaningfully compared. 

\begin{figure}[h]
    \centering
    \begin{subfigure}{0.45\linewidth}
        \includegraphics[width=\linewidth]{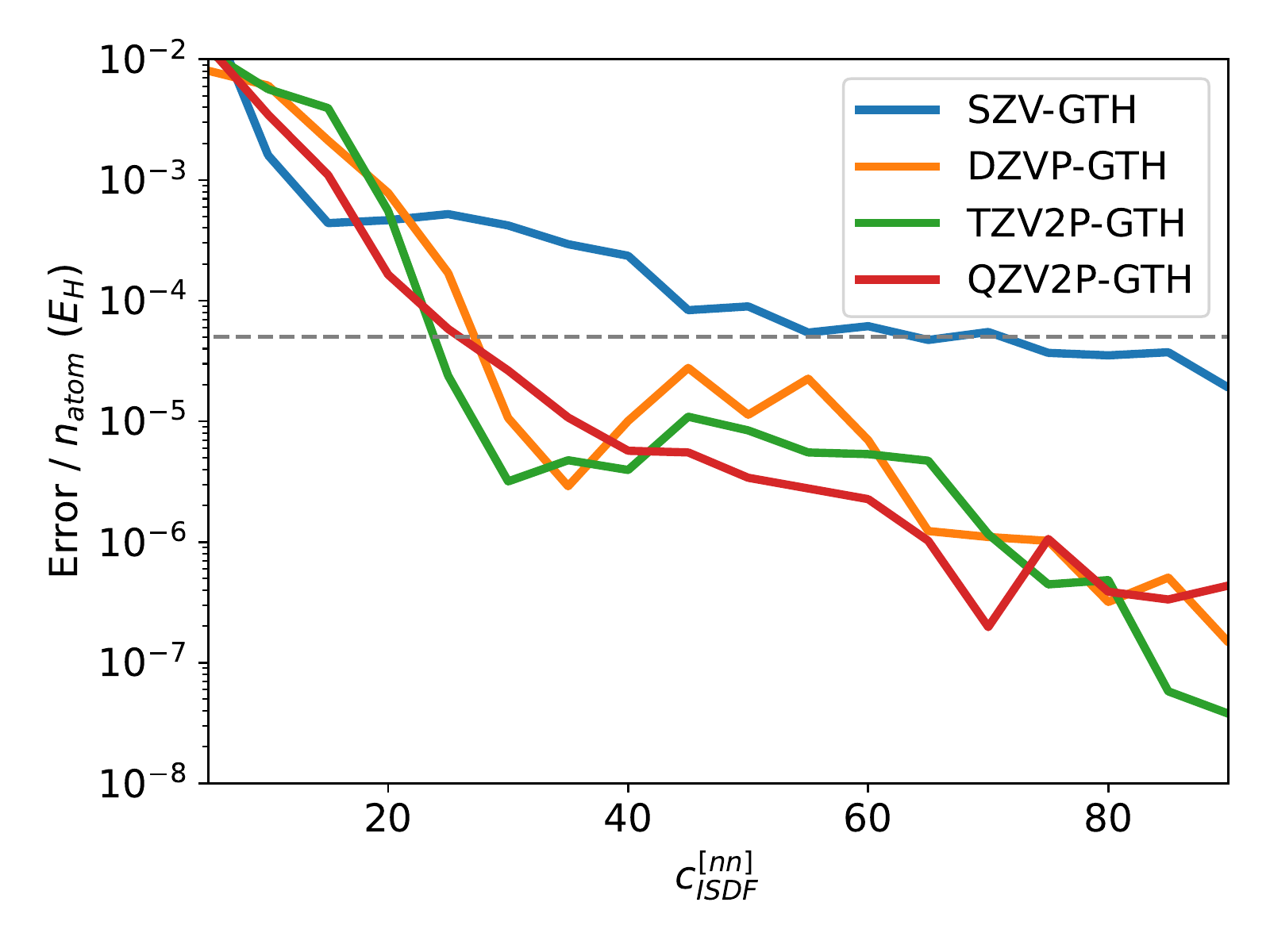}
        \caption{}
        \label{fig:err-vs-c-vs-basis_aa}
    \end{subfigure}
    \begin{subfigure}{0.45\linewidth}
        \includegraphics[width=\linewidth]{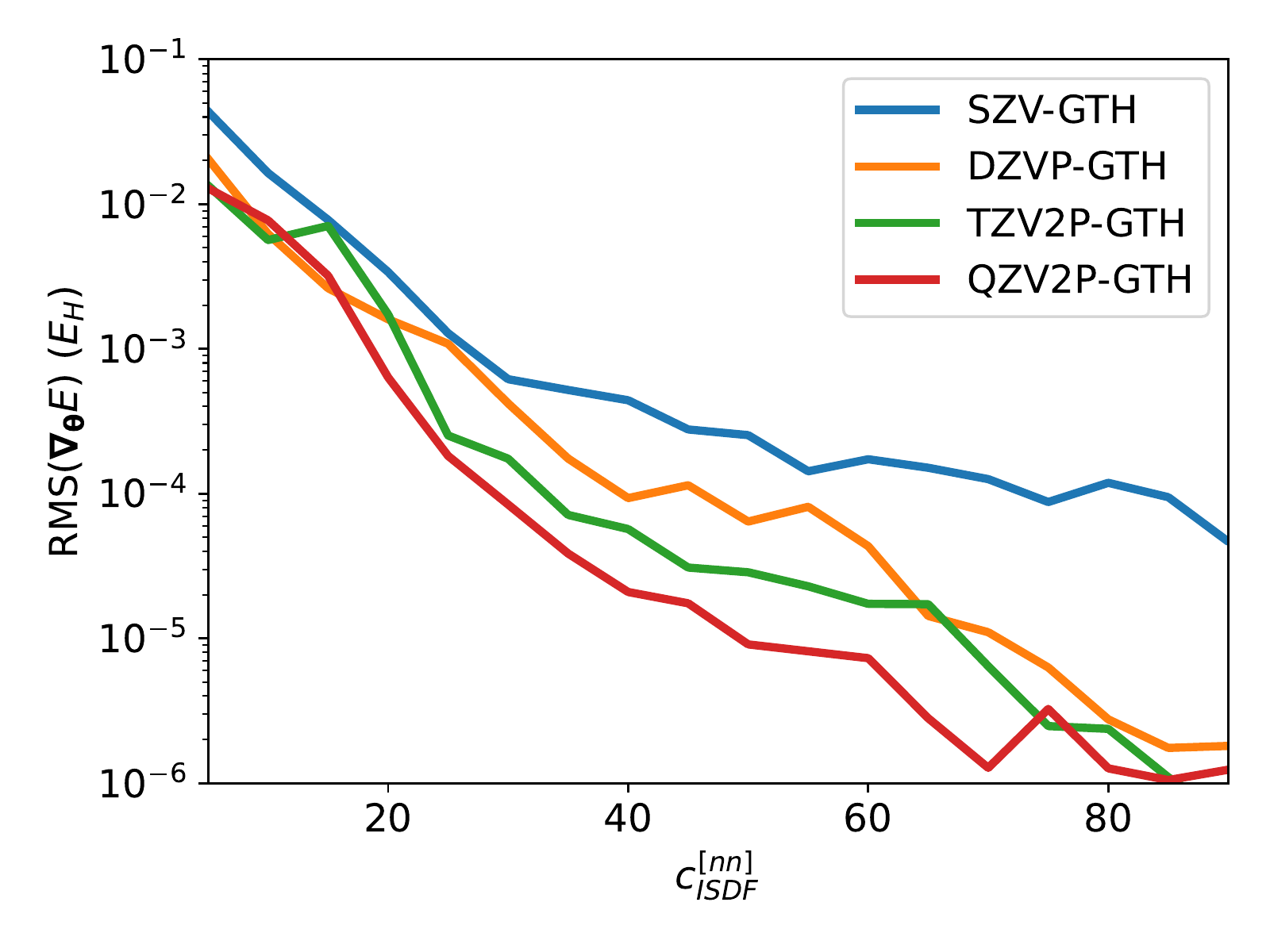}
        \caption{}
        \label{fig:grad-vs-c-vs-basis_aa}
    \end{subfigure}
    \\
    \begin{subfigure}{0.45\linewidth}
        \includegraphics[width=\linewidth]{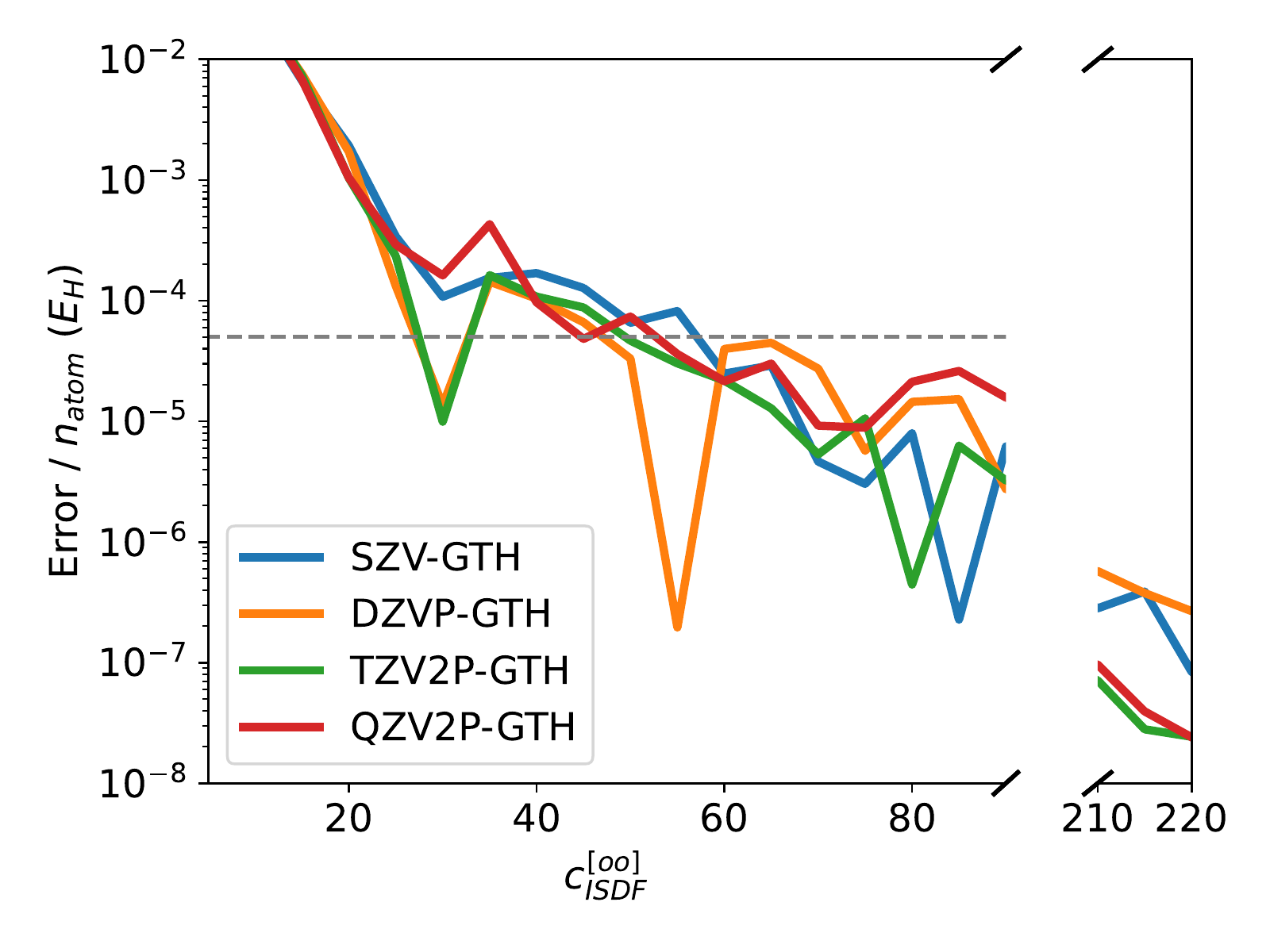}
        \caption{}
        \label{fig:err-vs-c-vs-basis_oo}
    \end{subfigure}
    \begin{subfigure}{0.45\linewidth}
        \includegraphics[width=\linewidth]{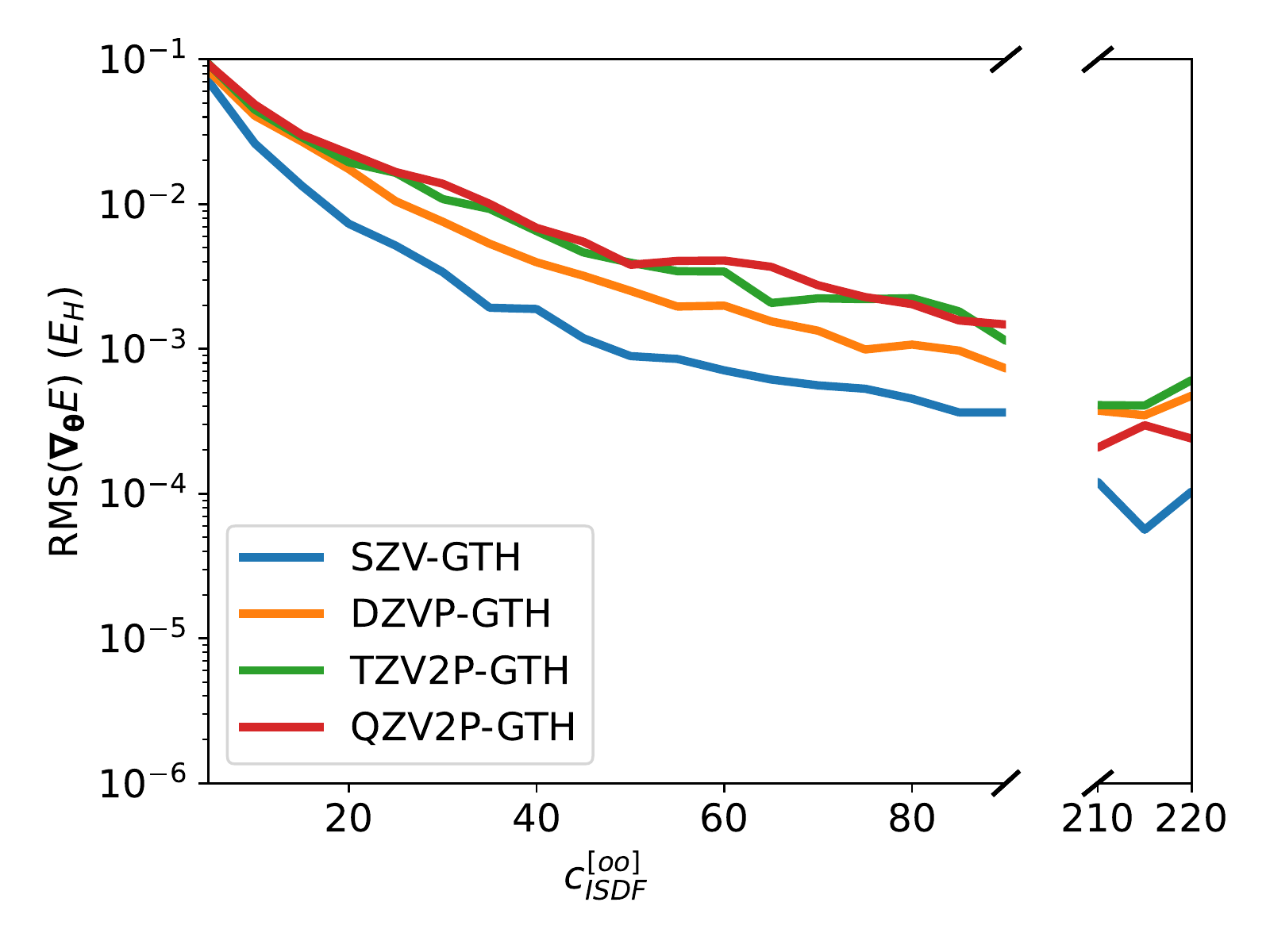}
        \caption{}
        \label{fig:grad-vs-c-vs-basis_oo}
    \end{subfigure}
    \caption{Error in the THC-AO-K HF energy (a), THC-AO-K orbital gradient (b), THC-oo-K HF energy (c), THC-oo-K orbital gradient (d), for diamond with a (2,2,2) $\mathbf k$-point mesh, given the converged density computed without the ISDF quadrature error. The numbers of ISDF interpolation points used for THC-AO-K and THC-oo-K are related to the parameters $c_{\text{ISDF}}^{[nn]}$ and $c_{\text{ISDF}}^{[oo]}$ via Eqs. \ref{eq:N_ISDF^nn} and \ref{eq:N_ISDF^oo} respectively.}
    \label{fig:egrad-vs-c-vs-basis_aa}
\end{figure}

We first investigate the accuracy of the THC energy computed for both algorithms. In Figure \ref{fig:err-vs-c-vs-basis_aa} we show the absolute error per atom in the THC energy computed for diamond (8 electrons total) with a (2,2,2) $\mathbf k$-point mesh given the converged SCF density for several different basis sets. We see that THC-AO-K can accurately represent the energy using $c_{\text{ISDF}}^{[nn]} = 25$ for all bases other than SZV-GTH. SZV-GTH is the \textit{smallest} basis and from Eq. \ref{eq:N_ISDF^nn}, we see that the same $c_{\text{ISDF}}^{[nn]}$ parameter gives rise to smaller numbers of ISDF points for smaller basis sets. The smallest useful $c_{\text{ISDF}}^{[nn]} = 25$ value will therefore depend somewhat on basis set size. Importantly, we see that the THC-AO-K energy is systematically improvable, so one may increase the number of points used if a more accurate energy is desired. We do observe a some fluctuations in the error as $c_{\text{ISDF}}^{[nn]}$ is increased - the error is not strictly monotonically decreasing; this is likely due to the randomness inherent to the CVT algorithm used to select ISDF points. 

In Figure \ref{fig:grad-vs-c-vs-basis_aa}, we show the root mean square of the orbital gradient for the same system. As we are using non-THC converged orbitals, we expect this value to be zero if there were no ISDF error. We see that at $c_{\text{ISDF}}^{[nn]} = 25$, the gradient still has a magnitude of $\sim 10^{-3}$ for some basis sets, suggesting that the THC-AO-K orbitals are slightly different than the converged non-THC algorithms, due to the ISDF factorization error. This is also systematically improvable through the $c_{\text{ISDF}}^{[nn]}$ parameter and will eventually converge to the exact solution once enough points are used. Consistent with the discussion of the energies, we see that typically (though not always due to ISDF point selection as already mentioned) a given value of $c_{\text{ISDF}}^{[nn]}$ leads to larger RMS gradients for the smaller basis sets.

We show similar plots for THC-oo-K in figure \ref{fig:err-vs-c-vs-basis_oo} and \ref{fig:grad-vs-c-vs-basis_oo}. We see that a larger value of $c_{\text{ISDF}}^{[oo]} = 50$ is needed to reach an acceptable error for this algorithm. It should, however, be pointed out that the number of ISDF points for this algorithm is determined by scaling $N_{\text{occ}}$ (Eq. \ref{eq:N_ISDF^oo}), and this leads to significantly fewer points than THC-AO-K despite the higher $c_\text{ISDF}$. Accordingly, this means that the number of ISDF points used in THC-oo-K for each of the basis sets tested is the same! THC-oo-K is, therefore, highly effective at compactly representing the HF exchange energy for large basis sets. On the other hand, looking at the gradient in figure \ref{fig:grad-vs-c-vs-basis_oo}, we see that the THC-oo-K orbital gradient is significantly larger than that of THC-AO-K and the convergence with increasing $c_{\text{ISDF}}^{[oo]}$ seems to be quite slow. Moreover, there seems to be a degree of basis set dependence as a larger basis set has many more degrees of freedom; the SZV-GTH basis has a much smaller error in the gradient, DZVP-GTH, TZV2P-GTH, and QZV2P-GTH all have roughly similar errors. This inaccuracy in the gradient could be due to fluctuations in the THC error over the entire energy surface and could potentially lead to SCF convergence issues. We will next investigate self-consistent energies obtained via THC-K algorithms to address this.

\begin{figure}
    \centering
    \begin{subfigure}{0.45\linewidth}
        \includegraphics[width=\linewidth]{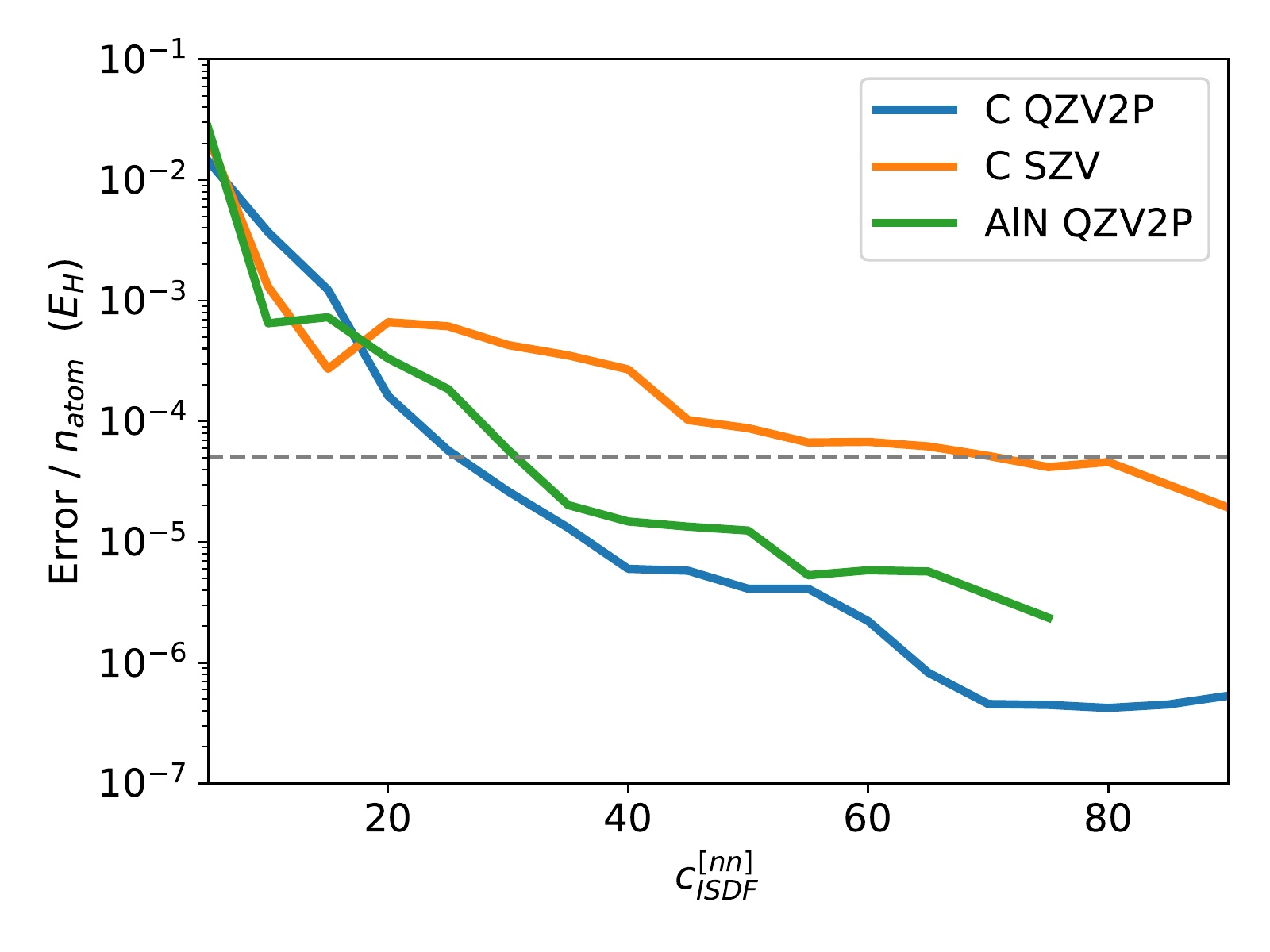}
        \caption{}
         \label{fig:err-vs-c_aa}
    \end{subfigure}
    \begin{subfigure}{0.45\linewidth}
        \includegraphics[width=\linewidth]{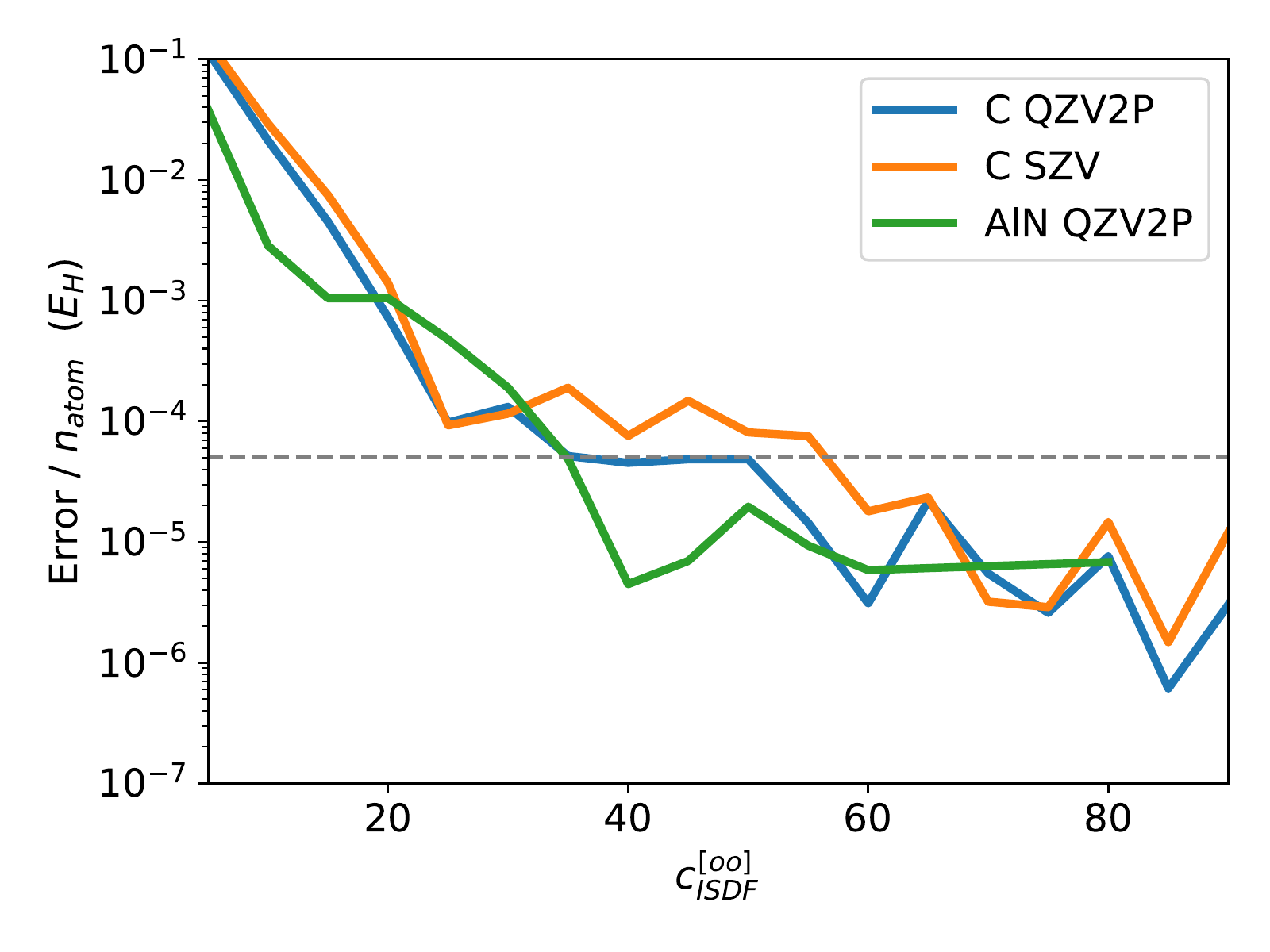}
        \caption{}
         \label{fig:err-vs-c_oo}
    \end{subfigure}
    \caption{Absolute Error per atom in the unit cell in the self-consistent HF energy using (a) THC-AO-K vs $c_\text{ISDF}^{nn}$ (see Eq. \ref{eq:N_ISDF^nn}) and (b) THC-oo-K vs. $c_\text{ISDF}^{oo}$ (see Eq. \ref{eq:N_ISDF^oo}) for both diamond and AlN using varying basis set sizes and a (2,2,2) $\mathbf k$-point mesh. The dashed line represents 50 $\mu E_H$ / atom accuracy.}
    \label{fig:err-vs-c}
\end{figure}

We next investigated the error in the HF energy for the converged density - i.e. we optimize orbitals using the THC algorithms. In figure \ref{fig:err-vs-c_aa}, we show the error in the HF self-consistent total energy using THC-AO-K as a function of $c_{\text{ISDF}}^{[nn]}$ for diamond in the QZV2P-GTH and SZV-GTH bases, as well as AlN (16 electrons total) in the GTH-QZV2P basis, all with a (2,2,2) $\mathbf k$-point mesh. We see that our suggested value of $c_{\text{ISDF}}^{[nn]} = 25$ holds up pretty well for the QZV2P-GTH systems, although diamond in the SZV-GTH basis does have a noticeably higher error in this case. A similar plot for THC-oo-K can be seen in figure \ref{fig:err-vs-c_oo}. We see that despite possible concerns arising from significant deviations in the orbital gradient when evaluated with exact converged density, the THC-oo-K algorithm also yields adequately accurate self-consistent solutions for these systems. Most encouragingly, the  $c_{\text{ISDF}}^{[oo]}$ parameter is evidently transferable across different systems and basis sets. 

\begin{figure}
    \centering
    \includegraphics[width=\linewidth]{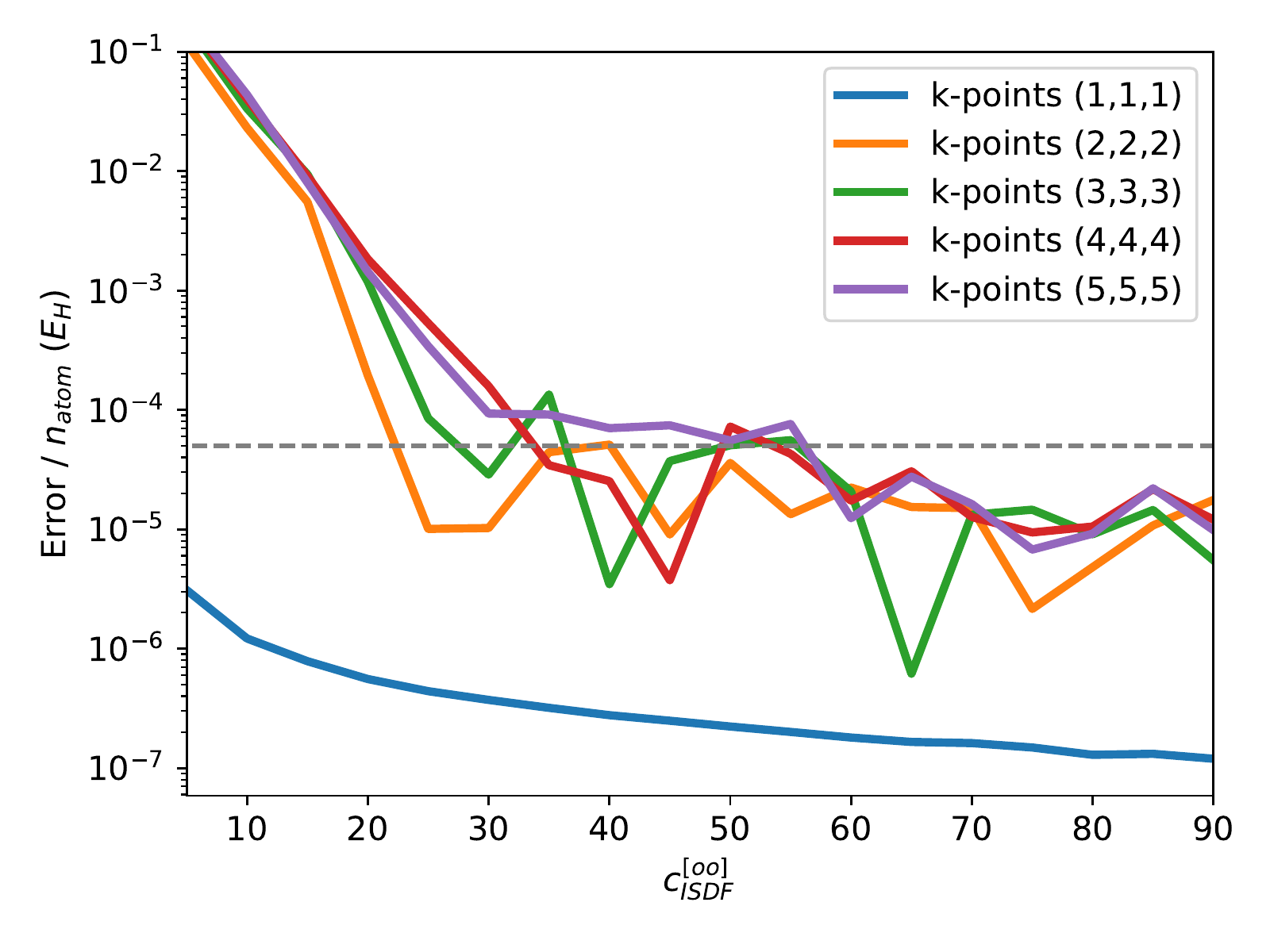}
    \caption{Absolute Error in the HF energy using THC-oo-K, as a function of $c_\text{ISDF}^{oo}$ for diamond in the QZV2P-GTH basis with several $\mathbf k$-point meshes.}
    \label{fig:err-vs-c-vs-k}
\end{figure}

We next investigate the accuracy as the $\mathbf k$-point mesh is increased. Our algorithms rely on the fact that the number of ISDF points required to yield an accurate solution will plateau quickly with the $\mathbf k$-mesh size. In Figure \ref{fig:err-vs-c-vs-k}, we show a plot of the absolute error in the self-consistent HF energy of diamond in the QZV2P-GTH basis computed with THC-oo-K for several different $\mathbf k$-point meshes. We see that there is a very steep increase in the number of ISDF points required when going from gamma point to a (2,2,2) $\mathbf k$-point mesh; evidently, it is much more difficult to represent products of occupied orbitals at multiple $\mathbf k$ points than at a single $\mathbf k$ point. However, increasing the mesh further has very little effect.  It therefore appears that the $c_{\text{ISDF}}^{[oo]} = 50$ value is adequate for all $\mathbf k$-point mesh sizes, although a significantly lower value may likely be sufficient for $\Gamma$-point calculations.

We therefore tentatively recommend starting with values  $c_{\text{ISDF}}^{[nn]} = 25$, $c_{\text{ISDF}}^{[oo]} = 50$. One may verify this is adequate on a case-by-case basis by increasing the parameters and ensuring little variation in overall results. We note that this study analyzed the absolute energies, which leaves out the effect of error cancellation in observables such as the lattice energy or band gaps. Error cancellation could lead to more leniency in the $c_{\text{ISDF}}$ parameters, necessitating fewer interpolation points. Additionally, we showed errors in the HF energy here, which includes 100$\%$ exact exchange. When using a hybrid density functional with a small fraction of exact exchange or short-range exact exchange, the THC errors will therefore be further scaled down. Indeed, this was observed in our molecular study.\cite{lee2019systematically} We therefore believe that our recommended coefficients are conservative.

\subsection{Computational scaling}
We found that both periodic THC-AO-K and THC-oo-K require more interpolation points than their molecular counterparts when using $\mathbf k$-point sampling. However, even molecular THC was seen to require too many interpolation points to be computationally competitive except for very large systems. We therefore investigate the $K$ build time for both THC-AO-K and THC-oo-K to see if ISDF-based algorithms are useful for small to medium-sized periodic systems. In the following calculations, we use $c_{\text{ISDF}}^{[nn]} = 25$ and $c_{\text{ISDF}}^{[oo]} = 55$. Additionally, we will compute the average THC-AO-K iteration time by assuming 10 iterations, as the majority of the computational cost of this algorithm is the $M$ build step which is performed only once, i.e. we add one-tenth of the $M$ build time into the per iteration cost. If more than 10 iterations are required (i.e., the initial guess was poor), THC-AO-K will be more favorable than shown here, and vice versa.

\begin{figure}[h!]
    \centering
    \begin{subfigure}{0.4\linewidth}
        \includegraphics[width=\textwidth]{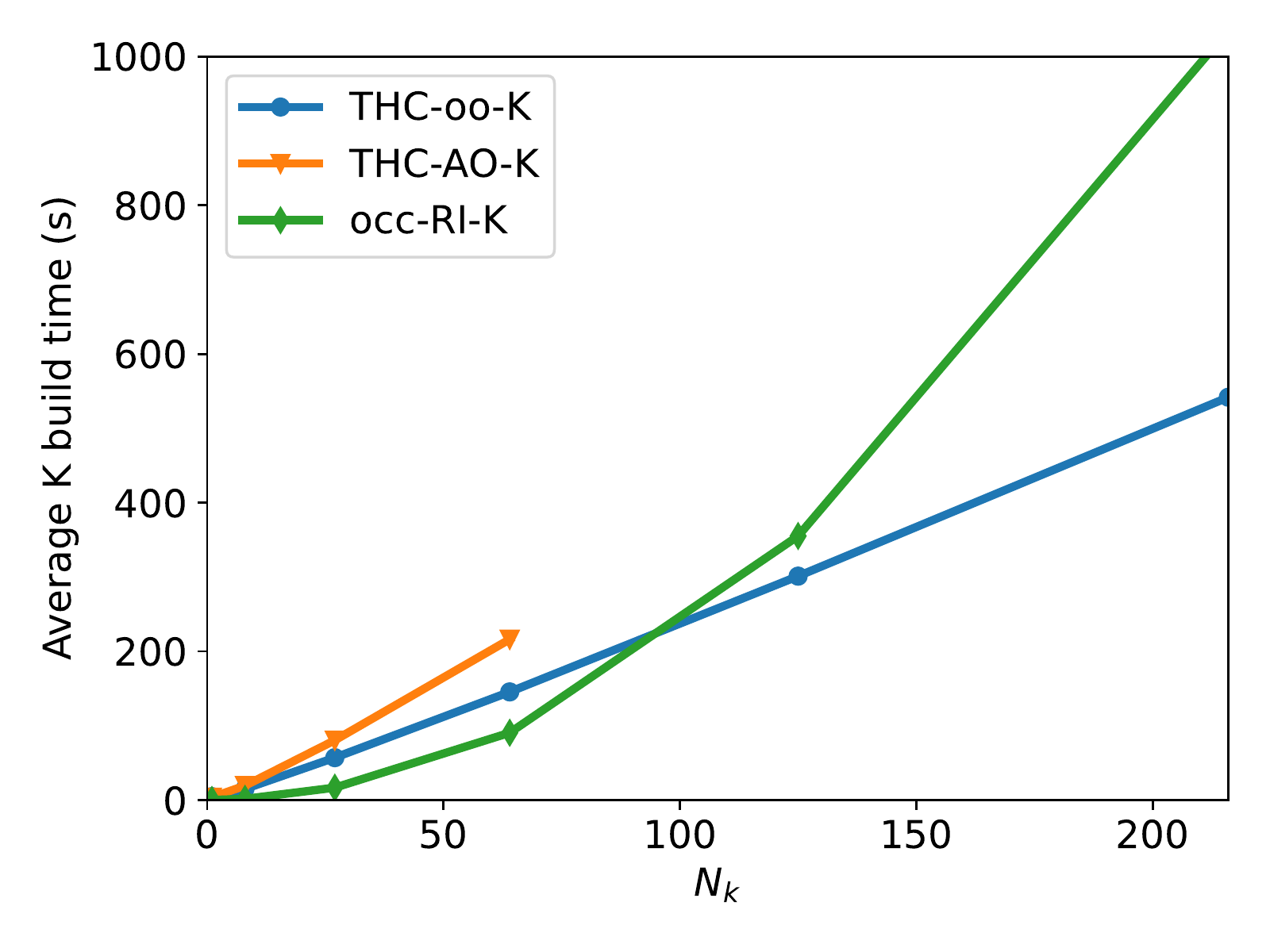}
        \caption{}
        \label{fig:t-vs-kpt}
    \end{subfigure}
    \begin{subfigure}{0.4\linewidth}
        \includegraphics[width=\textwidth]{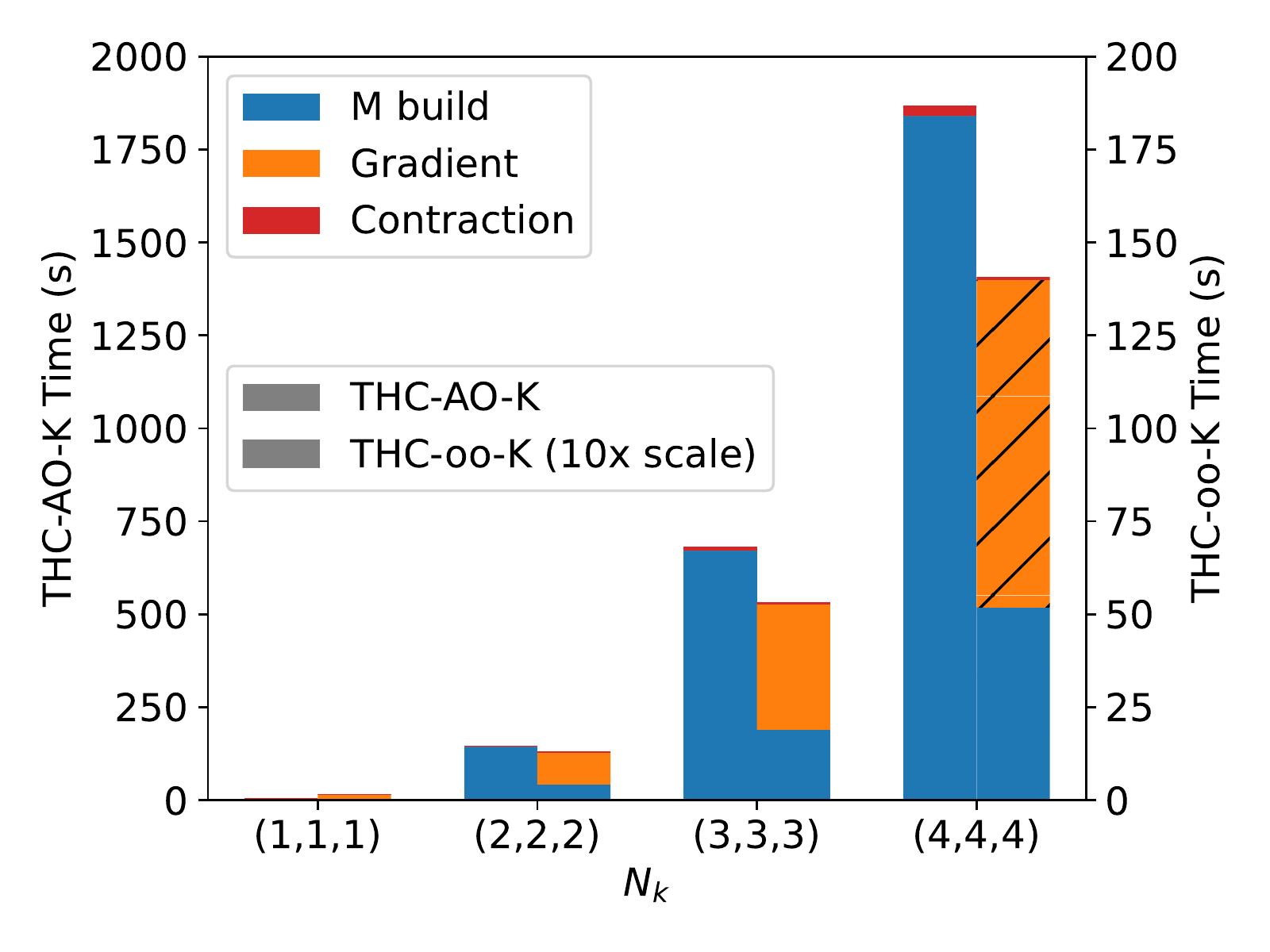}
        \caption{}
        \label{fig:tbreakdown}
    \end{subfigure}
    \caption{(a) Average $K$ build time per cycle and (b) breakdown of the THC initial cycle timings vs the number of $\mathbf k$-points for AlN in the QZV2P-GTH basis.}
    \label{fig:k-timings}
\end{figure}

In \cref{fig:t-vs-kpt}, we plot the $K$ build time for AlN in the QZV2P-GTH basis as a function of the number of $\mathbf k$-points for both THC-AO-K and THC-oo-K. We see two important conclusions from this plot. First, as anticipated, the THC-K algorithms are linear with the number of $\mathbf k$-points due to the FFT convolution. By contrast, occ-RI-K exhibits quadratic scaling with number of $\mathbf k$-points. Second, the THC-K algorithms have a large prefactor which causes occ-RI-K to be the more efficient approach for small $\mathbf k$-point meshes. The linear $\mathbf k$-point scaling makes THC-oo-K extremely effective at reducing compute time for large $\mathbf k$-point calculations, and the algorithm becomes more efficient than occ-RI-K between a (4,4,4) and (5,5,5) $\mathbf k$-point mesh. Due to the large basis set and therefore large $N_{\text{ISDF}}^{[nn]}$, the THC-AO-K algorithm is significantly slower than both other algorithms, and could not be run for $\mathbf k$-point meshes larger than (4,4,4) due to memory requirements.

We present a breakdown of the THC-AO-K and THC-oo-K first cycle computation time (the full THC-AO-K $\mathbf M$ build time is included) for the AlN system in \cref{fig:tbreakdown}. We see that the THC-AO-K time is dominated by the initial single $\mathbf M$ build, making the per iteration cost to contract of $\mathbf{M}$ to form $\mathbf K$ according to \cref{eq:kpointthck} negligible. The THC-oo-K $\mathbf M$ is significantly cheaper (note the 10x scale difference) to form, but the computation of the orbital gradient adds an extra factor of 2-3 to the per-iteration compute cost.

\begin{figure}[h!]
    \centering
    \includegraphics[width=\textwidth]{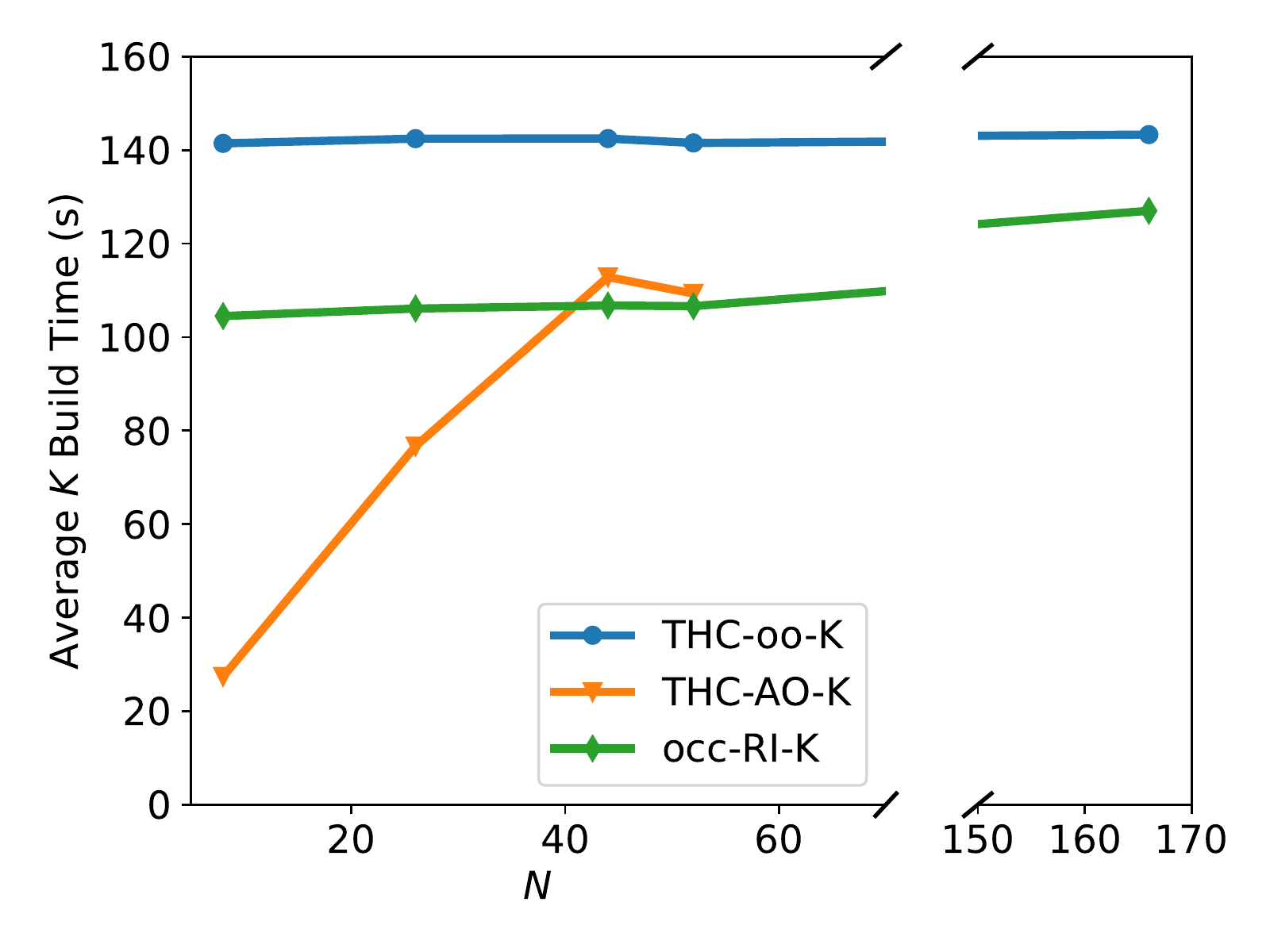}
    \caption{Average $K$ build time per cycle vs the number of basis functions for diamond with a (7,7,7) $\mathbf{k}$-point mesh.}
    \label{fig:t-vs-n}
\end{figure}

We additionally investigated the computational scaling of THC-K with basis set size. We previously saw the occ-RI-K algorithm was largely independent of basis set size, making it very effective for large basis calculations (with small numbers of $\mathbf k$-points). We expect THC-oo-K to have this same property, as $N_{\text{ISDF}}$ is independent of the basis size, shown by fig. \ref{fig:err-vs-c_oo}. We, therefore, expect THC-oo-K to become more efficient than THC-AO-K as the basis set size increases. In figure \ref{fig:t-vs-n}, we show the average $K$ matrix build time as a function of the basis set size for the diamond system with a (7,7,7) $\mathbf k$-mesh. THC-AO-K is more efficient than THC-oo-K and occ-RI-K for the smallest two basis sets used, SZV-GTH and DZVP-GTH; the bottleneck of this algorithm is precomputing the $M$ matrix, which is fairly small for smaller basis set sizes. occ-RI-K is slightly better than THC-AO-K for TZV2P-GTH and QZV2P-GTH as it is independent of basis size. THC-oo-K evidently does not become competitive with THC-AO-K until the very large unc-def2-QZVP-GTH basis set is used. 

\begin{figure}
    \centering
    \includegraphics[width=\textwidth]{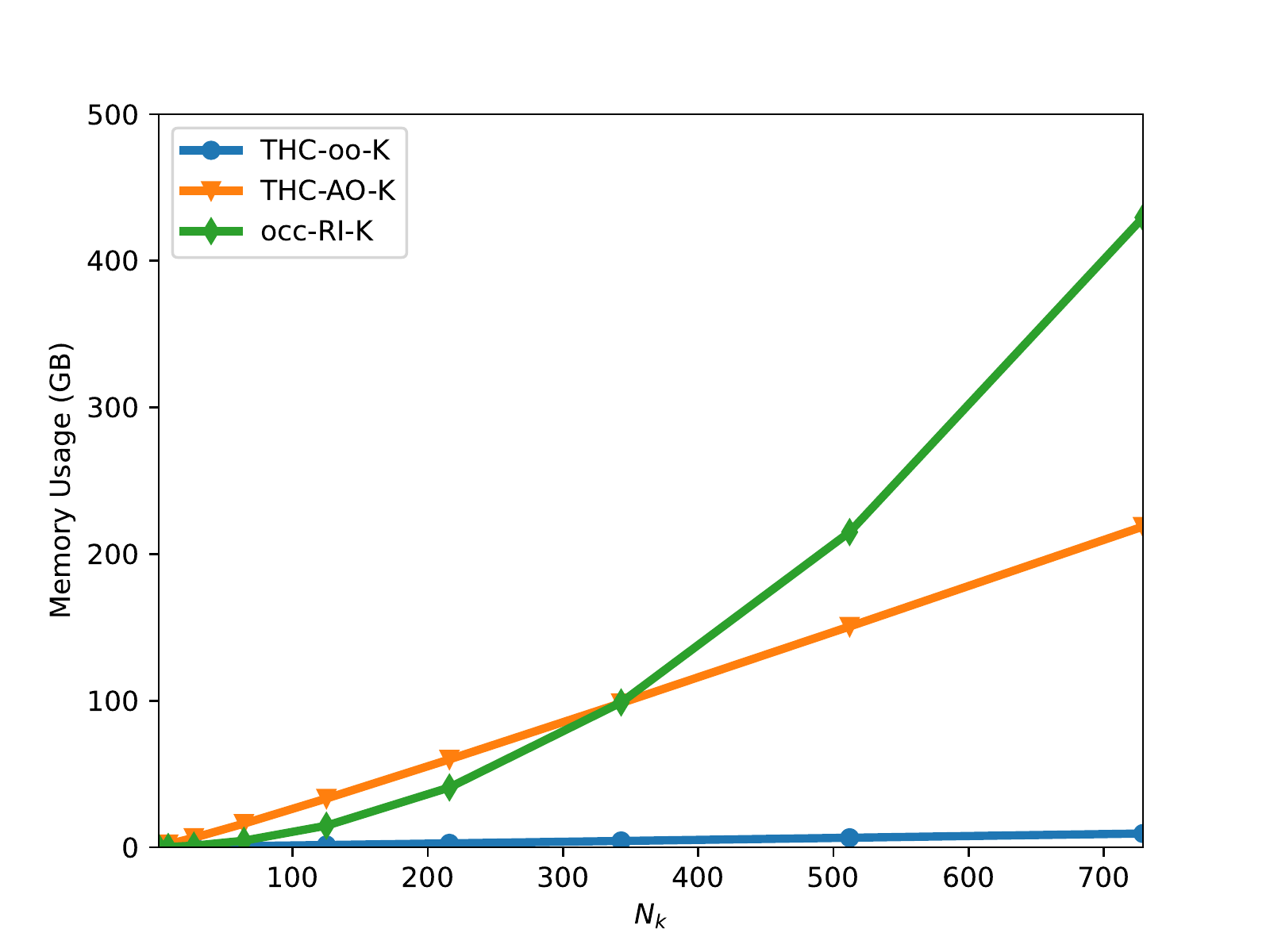}
    \caption{Memory  usage for the occ-RI-K, THC-AO-K, and THC-oo-K algorithms as a function of the number of $\mathbf k$-points for the diamond system using the QZV2P-GTH basis.}
    \label{fig:mem}
\end{figure}

Finally, we note that the THC algorithms can offer significantly reduced memory usage compared to occ-RI-K. For both THC algorithms, we store only $\xi_P(\mathbf{r})$ and $M_{PQ}^{\mathbf{q}}$, i.e. no quantities that scale with both the grid and number of $\mathbf k$-points. This is in contrast to the occ-RI-K algorithm which stores the basis functions (linear in $N_k$) and coulomb kernel (quadratic in $N_k$) on the grid by our default in Q-Chem. Using an integral direct approach to occ-RI-K will eliminate this memory requirement\cite{lee2022faster} but results in much slower performance, making it less competitive with the THC algorithms. We highlight this in figure \ref{fig:mem} by plotting the memory requirements for each algorithm for the diamond system in the QZV2P-GTH basis as a function of the number of $\mathbf k$-points. We see that the large number of ISDF points required for THC-AO-K leads to larger memory consumption than occ-RI-K for $\mathbf k$-point meshes smaller than (6,6,6), however, the linear scaling in $N_k$ eventually wins out at $\mathbf k$-point meshes larger than this. THC-oo-K offers a huge reduction in memory usage over THC-AO-K due to a large reduction in the number of ISDF points. For this diamond QZV2P-GTH example, with a (9,9,9) $\mathbf k$-point mesh, THC-oo-K requires only 9 GB compared to 430 GB for occ-RI-K!

We can now define specific use cases for each algorithm studied here. For small $\mathbf k$-point meshes, we recommend using the occ-RI-K algorithm as it will be faster than all THC algorithms. For large $\mathbf k$-point meshes we recommend using THC-AO-K if using a small basis set and THC-oo-K for large basis sets. Across all system sizes and $\mathbf k$-point mesh sizes, if the suggested algorithm requires too much memory, we recommend using THC-oo-K instead. It is particularly well-suited for implementation on memory-constrained computing devices, such as graphical processing units.

\subsection{Molecular Crystals}
Finally, we perform an illustrative computation of the cohesive energy of the benzene crystal as a rigorous test of the limits of THC-K. Accurate theoretical estimates of this value are available via high-order wavefunction theory calculations performed via fragment-based approaches \cite{yang2014ab,ringer2008first,kennedy2014communication}. Periodic calculations have thus far been fairly limited as HF and pure density functionals are known to fail for binding purposes \cite{civalleri2007ab} and higher order wavefunction theories are very expensive for periodic studies of benzene. Local density functionals have been shown to yield reasonable results when dispersion corrections are utilized\cite{moellmann2014dft}, and recent studies using periodic MP2 have been performed \cite{bintrim2022integral, neumann2005energy}. However, these periodic calculations are generally limited to small $\mathbf k$-point meshes or small basis sets, or both due to computational constraints. Only a few hybrid functionals have been tested for benzene\cite{civalleri2008b3lyp,loboda2018towards,price2023xdm,moellmann2014dft,cutini2016assessment}  due to the computational difficulty of obtaining exact exchange energies in the thermodynamic limit. We, therefore, decided to utilize the THC-oo-K algorithm to benchmark hybrid functional for cohesive energies in order to compare with theoretical best estimates.

We investigated the dispersion-corrected pure functional B97M-rV\cite{mardirossian2015mapping, mardirossian2017use} and the hybrid functionals PBE0-D3\cite{adamo1999toward, grimme2010consistent}, MN15\cite{haoyu2016mn15}, M06-2X\cite{zhao2008m06}, SCAN0-D3\cite{hui2016scan},  $\omega$B97X-rV,\cite{mardirossian2014omegab97x} and $\omega$B97M-rV \cite{mardirossian2016omega}. We attempted to converge our results to both the complete basis set (CBS) limit and the TDL, including counterpoise corrections (see Sec. \ref{compute-details}). We cover hybrid functionals utilizing both the D3 and rVV10 dispersion corrections and empirical functionals parameterized to include dispersion. We note that MN15 and M06-2X are not dispersion-corrected but their performance on molecular data sets of non-covalent interaction was found to be quite accurate.\cite{Mardirossian:2017b} The cohesive energies computed for varying sizes of $\mathbf k$-point meshes can be seen in table \ref{tab:benzene}. We performed our calculations using the QZV2P-GTH basis. 

Compared to the theoretical best estimate (TBE) of -54.58 kJ/mol \cite{yang2014ab}, we see a great variance in the performance of DFT. As expected, the dispersion treatment is paramount to obtaining accurate, cohesive energies. The DFT-D3 treatment of dispersion yields quite accurate results, with the global hybrid functionals PBE0-D3 and SCAN0-D3 performing well, giving errors of 3.4 and 2.0 kJ/mol, respectively. MN15 and M06-2X, parameterized to implicitly treat dispersion, yield extremely different results; MN15 is the most accurate of the functionals tested with an error of 1.1 kJ/mol while M06-2X performs substantially worse than all other functionals, giving an error of 13 kJ/mol. Use of the rVV10 dispersion correction yields unsatisfactory results as well; the pure functional B97M-rV overbinds by 9.0 kJ/mol while the range separated hybrids $\omega$B97M-rV and $\omega$B97X-rV, which are among the most accurate functionals for molecular systems\cite{Mardirossian:2017b}, perform only slightly better with errors of 7.2 and 8.0 kJ/mol respectively. The consistent overbinding of these three VV10-based functionals suggests that rVV10 is the main source of their errors.

The trends seen in our data are in line with previous studies on dispersion corrections for benzene and the X23 dataset. DFT-D3 is known to perform quite well\cite{moellmann2014dft}, while the failure of rVV10 for molecular crystals has previously been attributed to the lack of screening (i.e., many-body dispersion) effects\cite{hermann2018electronic, stohr2019theory, fabiano2023seeking}. Notably, by comparing B97M-rV to $\omega$B97M-rV,  we find that the addition of exact exchange makes a minor improvement of 1 kJ/mol.

\begin{figure}
    \centering
    \includegraphics[width=\textwidth]{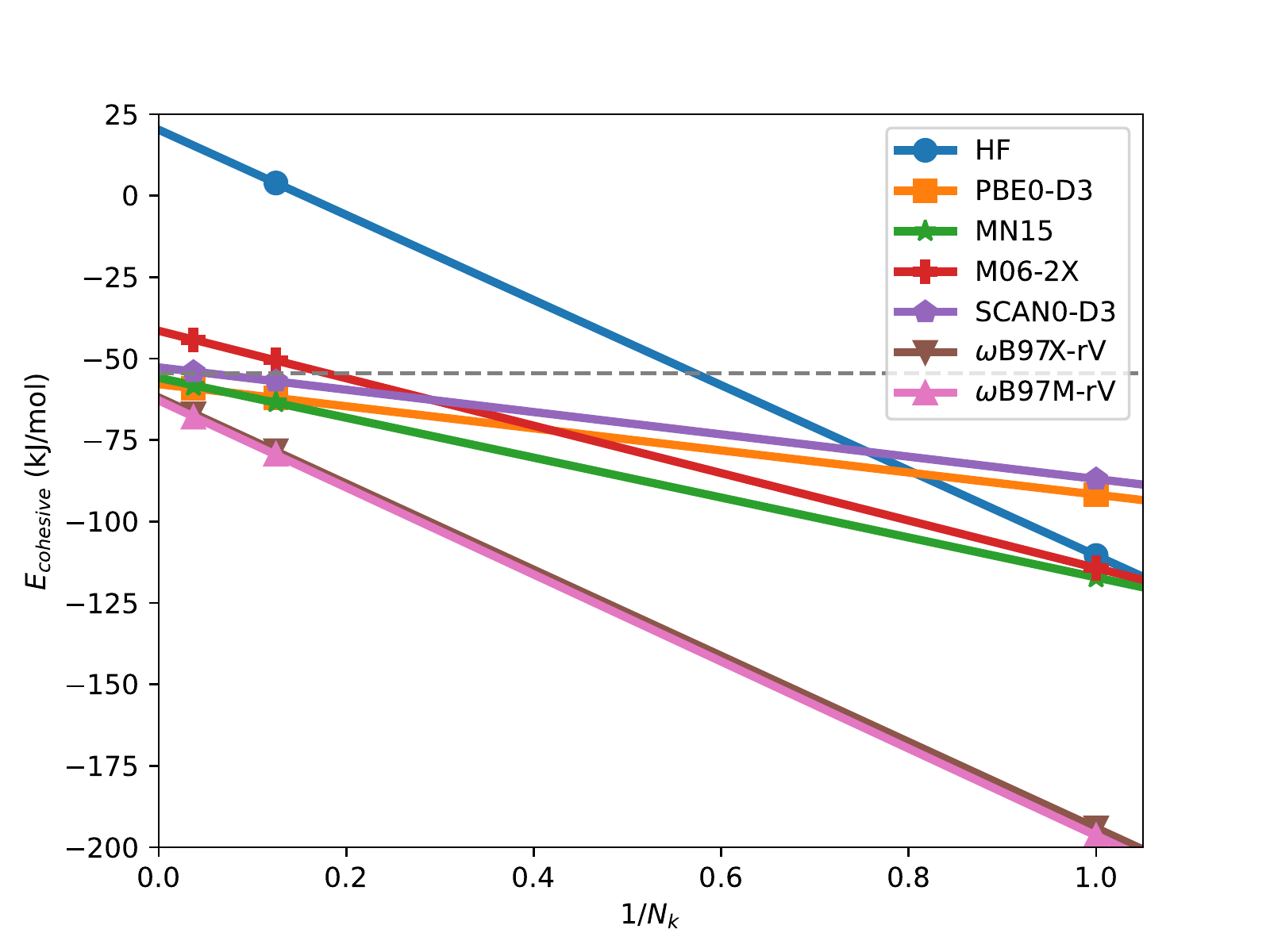}
    \caption{Cohesive energy of the benzene crystal as a function of $\mathbf k$-point sampling for several hybrid functionals. The dashed line indicates the reference value.}
    \label{fig:benzene}
\end{figure}

We see that with a (3,3,3) $\mathbf k$-point mesh, pure functionals are essentially converged, whereas the hybrid functionals are still up to $5$ kJ/mol off from the TDL. The convergence of the hybrid functionals is demonstrated in figure \ref{fig:benzene}, where the convergence rate is dependent on the fraction of exact exchange in the functional - highlighting the difficulty of periodic hybrid DFT. Nonetheless, the size extrapolation using $1/N_k$ works well for this system, making reliable hybrid functional calculations accessible.

\begin{table}[]
\begin{tabular}{|l|rrrrrrrr|}
\hline
$N_k$ &  B97M-rV &       HF & PBE0-D3 &    MN15 &  M06-2X & SCAN0-D3 & $\omega$B97X-rV & $\omega$B97M-rV \\
\hline
$1^3$ &    -10.97 &  -55.87  & -37.18  &  -62.59 & -59.70  &   -32.38 &         -139.54 & -141.48   \\
$2^3$ &    -8.29 &   58.48   &   -7.44 &   -8.81 &   4.09  &    -2.33 &          -23.70 &  -24.74   \\
$3^3$ &    -8.98 &           &   -4.59 &   -3.76 &  10.33  &     0.54 &          -12.42 &  -13.40   \\
\hline
TDL   &    -8.98 &   74.81  &   -3.38 &    -1.63 &   12.95 &     1.75 &           -7.68 &  -8.62    \\
\hline
\end{tabular}
\caption{Error in the benzene crystal cohesive energy in kJ/mol for several density functionals as a function of the $\mathbf k$-point mesh size.}
\label{tab:benzene}
\end{table}

We see that for the benzene crystal, including exact exchange can lead to  modest improvements in accuracy, but the treatment of dispersion is much more important. More in-depth analysis is required to evaluate general trends of hybrid functionals for periodic applications. Still, the discrepancy between molecular\cite{Mardirossian:2017b} and periodic cases suggests that further functional development may be beneficial for hybrids applied to periodic applications.

%
%
\section{Conclusions}
To summarize, we have presented an extension of the ISDF approximation for exact exchange for use with periodic $\mathbf k$-point calculations using GTOs. We additionally presented a new algorithm, THC-oo-K, by fitting only the product of occupied orbitals via the ISDF approach, which is all that is needed for the energy. While this means that errors in THC-oo-K orbital gradients (which involve occupied-virtual products) are larger than for the energy, the effect on the self-consistently optimized energy is still small. We have shown that these algorithms reduce the computational scaling for exact exchange to cubic in system size and linear with the number of $\mathbf k$-points; the THC-oo-K algorithm has the additional advantage of THC dimension scaling independently of basis set size. Initial investigation showed that these algorithms provide substantial computational savings for large $\mathbf k$-point meshes; THC-oo-K additionally provides computational savings over occ-RI-K at even medium-sized $\mathbf k$-point meshes and huge reductions in memory usage in all cases, at only a minor compromise in overall accuracy. We believe this cost reduction will make periodic studies using hybrid functionals more feasible, which we illustrated via a study of the benzene lattice energy, computed with the QZV2P-GTH basis up to a (3,3,3) $\mathbf k$-point mesh.

Further reduction in error may be obtained by fitting only one side of the two-electron integral tensor via ISDF, termed robust fit THC.\cite{Pierce2021Apr,sharma2022fast} This reduces the error in the energy to quadratic in the THC fit error rather than linear. We performed preliminary studies of this approach for the THC-AO-K algorithm. Still, we found that while the robust fit did reduce error substantially, it significantly increased memory requirements to an intractable degree due to the requirement of storing $V^q_P(r)$, the potential due to interpolating function $\xi_P(r)$ which increases the memory usage of the already expensive THC-AO-K by a factor of $N_k$; additionally, the number of expensive FFTs for the GPW algorithm is increased by a factor of $N_k$, significantly increasing the compute time. Nonetheless, a multi-node MPI implementation may increase the available memory enough to make this approach tractable.\cite{sharma2022fast} For the THC-oo-K algorithm, there is no need to store $V^q_P(r)$ each iteration, so it is possible to batch over this quantity and avoid storing the entire thing; the increase in the number of FFTs will however still lead to a significant increase in compute cost. A robust fit THC-oo-K algorithm is therefore an interesting direction for future development, but the trade-off in compute time may turn out to be unfavorable. 

All work presented here was used with the GPW algorithm, which only applies to pseudopotential calculations. A big advantage of Gaussian orbitals in PBC calculations compared to the more popular plane waves basis is the ability to model core orbitals. It would be interesting to investigate all-electron versions of these algorithms. This extension should be possible using the projector augmented-wave method (PAW) \cite{blochl1994projector}, density fitting \cite{sun2017gaussian, ye2021fast}, or similar approaches, by selecting ISDF points from Becke grids as is done for molecular codes.\cite{lee2019systematically}

Finally, we note that this approach offers a general framework for approximating the two-electron integrals and its usefulness is not limited to exact exchange. The same intermediates may be used to speed up the computation of matrix elements necessary for M\o ller-Plesset theory, coupled cluster theory, and other correlated wavefunction methods. This has already proved successful for molecular ISDF\cite{hohenstein2012communication, lee2019systematically,lu2017cubic}. The extension of this algorithm to correlated wavefunction theory would allow for much larger $\mathbf k$-point calculations.

\vspace{20pt}
\noindent \textbf{Acknowledgements: }
This work was supported by the Director, Office of Science, Office of Basic Energy Sciences, of the U.S. Department of Energy under Contract No. DE-AC02-05CH11231. 
A.R. acknowledges funding from the National Science Foundation under award number DGE 1752814. 
J.L. thanks David Reichman for his support.

\bibliography{everything}

\end{document}